\documentclass[aps,prb,twocolumn,groupedaddress,showpacs]{revtex4}

\usepackage{graphicx}% Include figure files
\usepackage{dcolumn}% Align table columns on decimal point
\usepackage{bm}% bold math
\usepackage{amsmath,amssymb}

\begin{document}

%\preprint{}

\title{Critical magnetization behaviors of the triangular and Kagome lattice 
quantum antiferromagnets}



\author{T\^oru Sakai,$^{1,2,3}$ and Hiroki Nakano$^{2}$ }
\affiliation{$^1$ {Japan Atomic Energy Agency, SPring-8, Sayo, Hyogo 679-5148, Japan.}\\
$^2$ {Graduate School of Material Science, University of Hyogo, Kouto 3-2-1, Kamigori, Ako-gun, Hyogo 678-1297 Japan.}\\
$^3$TRIP JST, Saitama 332-0012, Japan
}


\date{\today}

\begin{abstract}

We investigate the $S=1/2$ quantum spin antiferromagnets on the triangular 
and Kagome lattices in magnetic field, using the numerical exact 
diagonalization. Particularly we focus on an anomalous magnetization behavior
of each system at 1/3 of the saturation magnetization. 
The critical exponent analyses suggest that it is a conventional 
magnetization plateau on the triangular lattice, 
while an unconventional phenomenon, called the magnetization ramp, 
on the Kagome lattice. 

\end{abstract}


\maketitle

%\section{}
%\subsection{}
%\subsubsection{}

The $S=1/2$ triangular and Kagome\cite{brief_history} 
lattice antiferromagnets have attracted 
a lot of interests as typical frustrated systems. 
Most theoretical studies indicated that the former system has the three 
sublattice long-range order\cite{huse,bernu,capriotti}, 
while the latter is disordered in the 
ground state\cite{sachdev,miyashita,Lecheminant,Waldtmann,mila,hermele,singh,ran,jiang,Cepas,Sindzingre,vidal}. 
Experimental studies to observe a novel spin liquid phase 
have been accelarated since discoveries of several realistic 
materials; 
the organic compound $\kappa $-(BEDT-TTF)$_2$Cu$_2$(CN)$_3$ 
for the triangular lattice\cite{kanoda}, 
the herbertsmithite\cite{herbertsmithite_jacs,herbertsmithite_jpsj}, 
the volborthite\cite{volborthite_jpsj_let,volborthite_prl} 
and the vesignieite\cite{vesignieite_let} for the 
Kagome lattice. 
Since the quantum Monte Carlo simulation and the DMRG calculation 
are useless for these systems, 
the numerical exact diagonalization is one of the best 
numerical method for them. 
The numerical diagonalization studies suggested that 
both systems have the 1/3 magnetization plateau
\cite{Hida,Cabra,Honecker0,Honecker1,Honecker2}, 
although the classical spin systems have no plateau on both lattices 
in the ground state\cite{kawamura,mike}. 
%
% addition 2
%
(The thermal or quantum fluctuations induce a plateau in the 
semiclassical case, because 1/3 is just a critical point between 
two dirrerent spin structures.) 
In our recent numerical diagonalization study on the $S=1/2$ Kagome lattice 
antiferromagnet up to $N=36$, 
the calculated field derivatives reveal 
an anomalous behavior at 1/3 of the saturation magnetization\cite{nakano}. 
Namely, the field derivative is diversing at the low-field side of 
the critical field $H_c$, while almost zero at the high-field side. 
This critical behavior is quite different from conventional magnetization 
plateaux in two-dimensional systems where the field derivative 
is finite at both sides of $H_c$. 
To distinguish such an anomalous property at the 1/3 magnetization 
of the Kagome lattice from conventional plateaux, 
we called it a ``magnetization ramp''. 
However, its mechanism is still an open problem. 
In this paper, to clarify such an unconventional behavior around the 
1/3 magnetization of the $S=1/2$ Kagome lattice antiferromagnet, 
comparing with the triangular one, we applied the numerical 
diagonalization for both systems up to 
$N=39$ which is the largest cluster at present.  
In addition we estimated the critical exponent $\delta$ by the 
finite-size scaling proposed by the previous work\cite{sakai1}, 
to investigate the quantum critical behevior more quantitatively.

\begin{figure}
\includegraphics[width=0.95\linewidth,angle=0]{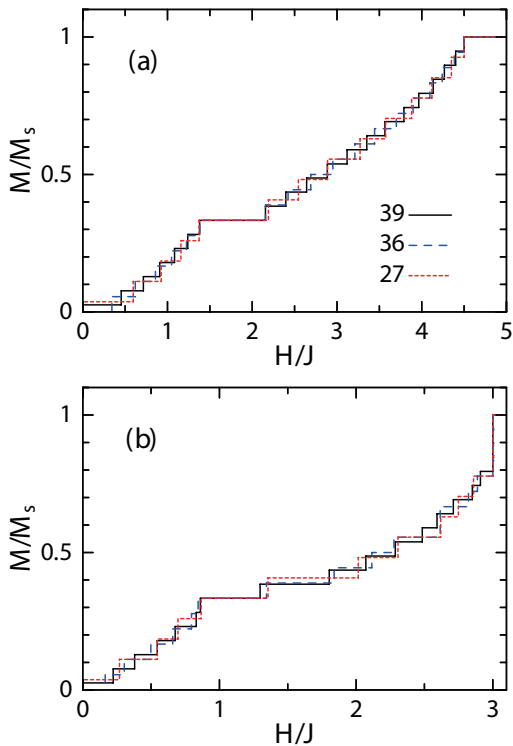}%
\caption{\label{mag} 
Magnetization curves of the (a) triangular and (b) 
Kagome lattice antifferomagnets for $N=$27, 36, 39. 
}
\end{figure}

\begin{figure}
\includegraphics[width=0.85\linewidth,angle=0]{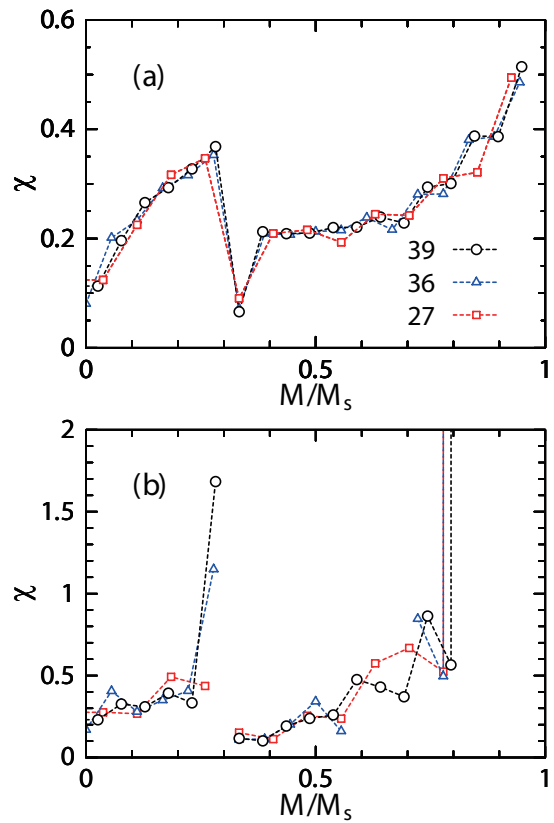}%
\caption{\label{chi} 
Field derivatives of the (a) triangular and (b) 
Kagome lattice antifferomagnets for $N=$27, 36, 39. 
}
\end{figure}

%\section{Model}
Now we examine the magnetization process 
of the $S=1/2$ triangular and kagome lattice antiferromagnets. 
The Hamiltonian is given by 
\begin{eqnarray}
\label{ham}
&{\cal H}&={\cal H}_0+{\cal H}_Z, \\
&{\cal H}_0& = \sum _{\langle i,j \rangle} {\bf S}_i \cdot {\bf S}_j, 
\qquad  
{\cal H}_Z =-H\sum _j^N S_j^z, 
\end{eqnarray}
where $\langle i,j \rangle$ means all the nearest neighbor pairs 
on each lattice. 
Throughout we use the unit such that $g \mu _B =1$. 
For $N$-site systems, 
the lowest energy of ${\cal H}_0$ in the subspace where 
$\sum _j S_j^z=M$ 
(the macroscopic magnetization is 
$m=M/M_{\rm s}$, 
where $M_{\rm s}$ denotes the saturation of the magnetization, 
namely $M_{\rm s}=N S$ for the spin-$S$ system. ) 
is denoted as $E(N,M)$. 
We restrict us to the rhombic cluster under the periodic boundary 
condition to keep the 120$^{\circ}$ rotational symmetry 
for a systematic finite-size scaling. 
Using the numerical exact diagonalization, 
we have calculated 
all the values of $E(N,M)$ available for the rhombic 
clusters with $N=$9, 12, 21, 27, 36 and 39, to 
obtain the ground state magnetization curves. 
(The largest dimension of the $N=39$ system is 68,923,264,410. 
To treat such huge matrices in computers, we have carried out 
parallel calculations using the MPI-parallelized 
code which is originally developed in the previous work\cite{HN_Terai}.)
The magnetization curves of the (a) triangular and (b) Kagome lattice 
antiferromagnets are shown in Fig. \ref{mag} for $N=$27, 36 and 39. 
They indicate plateau-like behaviors at $m=1/3$ of both systems, 
but the Kagome lattice exhibits an anomalous feature; 
the step length increases with decreasing $H$ towards $m=1/3$, 
different from conventional magnetization plateau like the 
triangular lattice. 
In order to clarify a difference between the triangular and 
Kagome lattice, we calculated the field derivative $\chi$ 
defined the form
\begin{equation}
\chi^{-1} = \frac{E(N,M+1)-2E(N,M)+E(N,M+1)}{1/M_{\rm s}}. 
\label{chi_calc}
\end{equation}
The derivative $\chi$ of the (a) triangular and 
(b) Kagome lattice systems are 
shown in Fig. \ref{chi} for $N=$27, 36 and 39. 
The derivative $\chi$ of the triangular system is 
finite at both edges of the 1/3 plateau-like behavior, 
as well as conventional two-dimensional systems. 
In contrast, the Kagome system exhibits a quite different feature 
between the lower and higher field sides of $m=1/3$;  
$\chi$ is diverging at the lower side like a plateau in 
one-dimensional systems, while is very small (possibly zero) 
at the higher one. 
The present calculation for $N=39$ more strongly supports 
a ramp-like behavior predicted by our previous work. 

\begin{figure}
\includegraphics[width=0.95\linewidth,angle=-90]{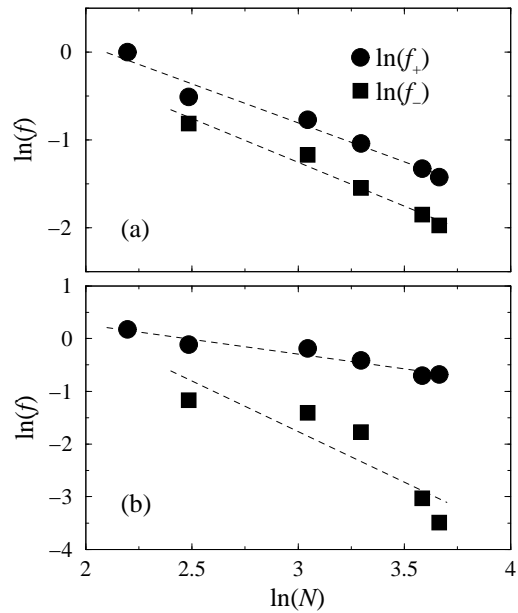}
\caption{\label{delta} 
$\ln(f)$ is plotted versus $\ln(N)$ for (a) the triangular and 
(b) Kagome lattice antiferromagnets, respecitively. 
Solid circles and squares are useful to estimate 
the critical exponents $\delta +$ and $\delta -$, respectively. 
 }
\end{figure}

The critical exponent $\delta$ defined by the form 
\begin{eqnarray}
\label{delta_def}
|m-m_c| \sim |H-H_c|^{1/{\delta}},
\end{eqnarray}
is an important index to specify the universality class of 
the field induced quantum phase transition. 
The previous theoretical works indicated $\delta=2$ for 
some typical one-dimensional gapped systems\cite{affleck,sakai2}, 
while $\delta=1$ for two-dimensional systems\cite{Katoh_Imada}. 
In order to investigate the quantum critical behavior at $m=1/3$ 
of the triangular and Kagome lattice antiferromagnets more 
quantitatively, we estimate $\delta$ by the finite-size scaling 
developed by the previous work\cite{sakai1}. 
Although it was proposed for one-dimensional systems, 
it can be easily generalized for two dimensions. 
We assume the asymptotic form of the size dependence of the energy as 
\begin{eqnarray}
\label{energy}
{1\over N}E(N,M)\sim \epsilon (m) +C(m) { 1\over {N^{\theta}}} 
\qquad (N\rightarrow \infty), 
\end{eqnarray}
where $\epsilon (m)$ is the bulk energy and the second term 
describes the leading size correction. 
We also assume that $C(m)$ is an analytic function of $m$. 
The lowest and highest magnetic field corresponding to $m=1/3$ 
in the thermodynamic limit are defined as $H_{c -}$ 
and $H_{c +}$, respectively, as the form 
\begin{eqnarray}
\label{lim}
E(N,{N\over 3}\pm 1)-E(N,{N\over 3}) \rightarrow \pm H_{c\pm} 
\quad (N\rightarrow \infty).
\end{eqnarray}
In order to consider the critical magnetization behaviors 
for $m < 1/3$ and $m> 1/3$ independently, we define the 
critical exponents $\delta _{c-}$ and $\delta _{c+}$ 
by the forms 
\begin{eqnarray}
\label{delta2}
|m-{1\over 3}| \sim |H-H_{c \pm}|^{1/{\delta _{\pm}}}.
%m-{1\over 3} \sim (H-H_{c+})^{1/{\delta _+}}, \\
%{1\over 3}-m \sim (H_{c-}-H)^{1/{\delta _-}}.
\end{eqnarray}
If we define the quantities $f_+(N)$ and $f_-(N)$ by the forms
\begin{eqnarray}
&&f_{\pm}(N) \nonumber \\ 
&&\equiv \pm[E(N,{N\over 3}\pm 2)+E(N,{N\over 3}) 
-2E(N,{N\over 3}\pm 1)],
\end{eqnarray}
the asymptotic forms of them are expected to be
\begin{eqnarray}
\label{f}
f_{\pm}(N)\sim {1\over {N^{\delta_{\pm}}}} 
+O({1\over{N^{\theta +1}}}) \qquad (N \rightarrow \infty ),  
\end{eqnarray}
as far as we assume the form (\ref{lim}). 
Thus the exponents $\delta _-$ and $\delta +$ can be estimated 
from the slope of the $\ln f_{\pm}$-$\ln N$ plot, 
respectively,  
under the condition $\theta > \delta_{\pm}-1$. 
In order to avoid an oscilation of the finite-size correction due to 
the cluster shape dependence, we just use the rhombic clusters 
under the periodic boundary condition 
with $N=$9, 12, 21, 27, 36, and 39. 
The plots of $\ln f_{\pm}$ versus $\ln N$ for the (a) triangular 
and (b) Kagome lattice antiferromagnets are shown in Fig. \ref{delta}. 
Figure \ref{delta} (a) suggests that 
the calculated points are well 
fitted to a line for each of $f_-$ and $f_+$ in the case of the 
triangular lattice. Thus applying the standard least square fitting 
to lines (Dashed and long-dashed lines are used to obtain $\delta_+$ 
and $\delta_-$, respectively.) for all the available system sizes; 
$N=$9, 12, 21, 27, 36, 39 ($N=9$ cannot be used for $\delta_-$), 
$\delta_-$ and $\delta_+$ are estimated as follows: 
\begin{eqnarray}
\delta _- =1.00 \pm 0.17, \qquad 
\delta _+ =0.89 \pm 0.15, \nonumber
\end{eqnarray}
for the triangular lattice. 
The errors are estimated from the deviation of points from the fitted lines. 
It would be reasonable to conclude $\delta_-=\delta_+=1$ at $m=1/3$ of the 
triangular lattice antiferromagnet, as expected for conventional 
magnetization plateau in two dimensions. 
On the other hand, Fig. \ref{delta} (b) indicates quite different 
feature of the Kagome lattice antiferromagnet. 
The same least square fitting yields the following estimates: 
\begin{eqnarray}
\delta _- =1.92 \pm 0.99, \qquad
\delta _+ =0.56 \pm 0.15, \nonumber
\end{eqnarray}
for the Kagome lattice antiferromagnet. 
Exponent $\delta_-$ has a large error 
because the line fitting is not good. 
It does not converge with respect to the system size well, but 
seems to still increasing with $N$. 
The same line fitting to the points for $N=$27, 36 and 39 
yields the estimation $\delta _-=4.59 \pm 0.25$. 
Thus we can just conclude $\delta _- \geq 2$ at most. 
It means that the diversing behavior of the field derivative 
at $H_{c-}$ is stronger than one-dimensional systems. 
It is led to two possibilities. 
One is a jump (a first-order transition) 
in the magnetization curve. 
A magnetization jump which also appears near the saturation 
was proved\cite{tsunetsugu}. 
The other is an anomalous continuous transition. 
A similar phenomenon was reported 
in the metal-insulator transition of the Hubbard chain 
with next-nearest neighbor hopping\cite{HN_YTakahash}. 
In comparison with $\delta_-$, $\delta _+$ is more 
conclusive, because the fitting error is much smaller. 
According to the above result of the line fitting, 
we conclude $\delta_+ =0.6 \pm 0.2$. 
Thus the field derivative $\chi$ should be zero at the 
higher field side of $H_{c+}$, because $\delta_+$ is 
smaller than unity. 
It also justifies a property of the magnetization ramp. 

\begin{figure}
\includegraphics[width=0.8\linewidth,angle=-90]{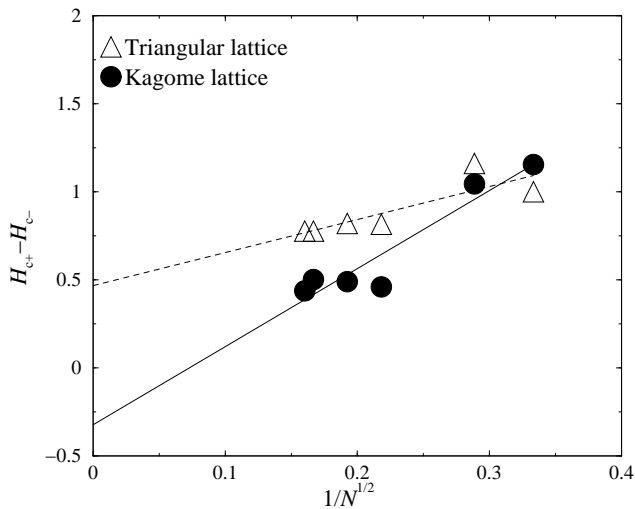}%
\caption{\label{plateau} 
Plateau width $H_{\rm c+}-H_{\rm c-}$ is plotted versus 
$1/N^{1/2}$. Open triangles and solid squares are for the 
triangular and Kagome lattice antiferromagnets, respectively. 
Fitted lines are used for the extrapolation to the 
thermodynamic limit. 
}
\end{figure}

Finally, we consider whether a flat part 
of the magnetization curve 
at $m=1/3$ exists or not for the triangular and Kagome lattice 
antiferromagnets. 
Namely, we examine whether each system has no plateau ($H_{c-}=H_{c+}$) 
or a finite plateau ($H_{c-}\not= H_{c+}$) at $m=1/3$ 
in the thermodynamic limit. 
We evaluate the length of the flat part $H_{c+}-H_{c-}$ 
corresponding to the 
plateau width of the finite-size clusters with $N=$9, 12, 21, 27, 36 
and 39 for both systems. 
If the system has a gapless excitation like a spin wave from some 
ordered states, the low-lying energy spectrum is expected to be 
proportional to the wave vector $k$ in the long wave lenght limit. 
Thus the excitation energy gap of the finite-size systems should 
have the asymptotic form $\sim 1/N^{1/2}$ in two-dimensional 
gapless systems. 
On the other hand, in gapped systems the gap is expected to 
converge to the thermodynamic limit with exponentially decaying 
(faster than $1/N^{1/2}$) 
finite-size correction, as the system size increases. 
Thus if the extrapolation 
by fitting the gap versus $1/N^{1/2}$ leads to a finite gap 
in the thermodynamic limit, it would be a strong evidence to 
confirm the gapped ground state. 
The length of a flat part $H_{c+}-H_{c-}$ is plotted versus $1/N^{1/2}$ 
in Fig. \ref{plateau}, where open triangles and solid circles are 
for the triangular and Kagome lattice antiferromagnets, respectively. 
The least square fitting to a line leads to the following results: 
$H_{c+}-H_{c-}=0.47 \pm 0.28$ for the triangular lattice and 
$H_{c+}-H_{c-}=-0.32 \pm 0.35$ for the Kagome lattice. 
Obviously we can conclude that the triangular lattice antiferromagnet 
has the 1/3 magnetization plateau. 
In contrast, the result for the Kagome lattice suggests that 
it possibly has a single critical field $H_c=H_{c-}=H_{c+}$. 
However, it is difficult to exclude a finite magnetization plateau, 
because of a large error of the extrapolation. 
%
% addition1
%
Note that any other plateaux are difficult to investigate by the 
present method, because fewer system sizes can be available for 
$m\not= 1/3$. 

In the recent magnetization 
measurement\cite{volborthite_jpsj_let} on a candidate 
of the Kagome lattice antiferromagnet Volborthite 
several step-like behaviors were observed, 
but it has not reached $m=1/3$ yet. 
The same measurement is sitll going on to observe an anomaly 
at $m=1/3$, which is expected to be about 60T. 
It would be interesitng to detect some unconventional features. 

In summary, we have investigated 
critical magnetization behaviors at $m=1/3$ 
for the $S=1/2$ triangular and Kagome lattice quantum antiferromagnets, 
using the numerical exact diagonalization of rhombic clusters up to $N=39$. 
The triangular lattice is revealed to have the critical exponets 
$\delta_-=\delta_+=1$ and a finite plateau, 
which are consistent with a conventional magnetization plateau 
in two-dimensional systems. 
On the other hand, 
the Kagome lattice is revealed to exhibit unconventional critical 
properties; $\delta_- < 1 < \delta _+$, namely the field derivative 
$\chi$ is diverging at the lower field side, while zero at the higher 
one of a possibly single critical field $H_c=H_{c-}=H_{c_+}$. 
The conclusion supports the magnetization ramp behavior at $m=1/3$ 
of the Kagome lattice antiferromagnet.

\begin{acknowledgments}

We wish to thank 
Profs.~K.~Hida 
and  ~F.~Mila 
for fruitful discussions. 
This work was partly supported by Grants-in-Aid 
(No.~20340096 and No.~22014012) 
for Scientific Research and Priority Areas 
``Novel States of Matter Induced by Frustration'' 
from the Ministry of Education, Culture, Sports, Science 
and Technology of Japan. 
Non-hybrid thread-parallel calculations 
in numerical diagonalizations were based 
on the TITPACK ver.2 coded by H. Nishimori. 
A part of the computations was performed using 
facilities of 
the Information Technology Center, Nagoya University, 
Department of Simulation Science, 
National Institute for Fusion Science, 
and the Supercomputer Center, 
Institute for Solid State Physics, University of Tokyo. 
\end{acknowledgments}

%\bibliography{lamdaBETS.bib}

\end{document}